\documentstyle[12pt,epsf]{article}
\newcommand{\RR}{\mbox{${\rm \:  R\!\!\!\! I
\;\;}$}} 

\newcommand{\qed}{\hfill $\Box$ \vskip 2ex}
\newcommand{\be}[1]{\begin{equation}\label{#1}}
\newcommand{\ee}{\end{equation}}
\newcommand{\vs}{\vspace{0.25cm}}

\begin{document}

\title{UNIFORM FINITE GENERATION OF COMPACT LIE GROUPS AND UNIVERSAL QUANTUM GATES}

\author{Domenico D'Alessandro \\ 
Department of Mathematics\\
Iowa State University \\
Ames, IA 50011,  USA\\
Tel. (+1) 515 294 8130\\
email: daless@iastate.edu}

\maketitle

\begin{abstract}
Consider a compact connected 
Lie group $G$ and the corresponding 
Lie algebra $\cal L$. Let $\{X_1,...,X_m\}$ 
be a set of generators for the Lie algebra 
$\cal L$. We prove that $G$ is uniformly finitely 
generated by $\{X_1,...,X_m\}$.  This means 
that every element $K \in G$ can be expressed as 
$K=e^{Xt_1}e^{Xt_2} \cdot \cdot \cdot e^{Xt_l}$, 
where the indeterminates 
$X$ are in the set $\{ X_1,...,X_m \}$, $t_i \in \RR$, $i=1,...,l$, 
 and the number $l$ 
is uniformly bounded. This extends a previous result by 
F. Lowenthal  
in that we do not require the connected one dimensional Lie 
subgroups corresponding to the $X_i$, $i=1,...,m$, to be
compact.

We discuss the consequence of this result to the question 
of universality of quantum gates in quantum computing.

\end{abstract}

\vs


\section{Introduction}

Let $G$ be a connected Lie group and $\cal L$ the corresponding 
Lie algebra and  let $\{X_1,...,X_m\}$ 
be a set of generators for $\cal L$. 
In their classical paper on controllability 
of systems on Lie groups  \cite{suss} 
 V. Jurdjevic and H. Sussmann proved that 
any element $K$ in $G$ can be written as 
\be{key}
K=e^{Xt_1}e^{Xt_2} \cdot \cdot \cdot e^{Xt_l}, 
\ee  
where the indeterminates  $X$'s are in the set  
$\{X_1,...,X_m\}$ and $t_1,...,t_l\in \RR$
(see Lemma 6.2 in \cite{suss}). The number $l$ called 
{\it order of generation} depends on 
$K$. F. Lowenthal called the Lie group 
$G$ {\it uniformly finitely generated} by 
$\{ X_1,...,X_m\}$ if every element 
$K \in G$ can be written in the form (\ref{key}) with $l$ 
uniformly bounded and described a number of Lie groups 
that are uniformly finitely  generated by 
every set of generators 
of the corresponding Lie algebra 
(see \cite{KL1}, \cite{KL2}, \cite{KL3} and
references therein). 
F. Lowenthal also showed in 
\cite{lowen} that, if $G$ is compact and 
the one dimensional connected 
Lie subgroups of $G$ corresponding 
to $\{X_1,X_2,...,X_m\}$ are also compact,  
 then the Lie group $G$ is uniformly finitely generated by
$\{X_1,X_2,...,X_m\}$. 
The aim of this note is to 
extend this result by relaxing the assumption of 
compactness of the one dimensional connected 
Lie groups corresponding to $\{X_1,...,X_m\}$. 
We refer to \cite{CSL1} \cite{CSL2} \cite{CSL3} 
for more discussion and results on the uniform 
generation problem.

\vs

\section{Uniform Finite Generation of Compact Lie Groups}

Consider the set of linearly independent generators of  $\cal L$,   
${\cal S}:= \{X_1,...,X_m\}$, we have the following lemma.

\vs

\noindent{\bf Lemma 1} {\it There exists a neighborhood $N$ 
of the identity 
in $G$ such that every element $K$ in  $N$ can be written as 
\be{writtenas}
K:=e^{Xt_1}e^{Xt_2}\cdot \cdot \cdot e^{Xt_l}.  
\ee 
The indeterminates $X$ 
are in the set $\cal S$, 
$t_j \in \RR$, $j=1,...,l$, and 
\be{bound}
l \leq n + 2\sum_{k=1}^{n-m}r_k.  
\ee
The sequence  $r_k$, $k=1,...,n-m$, is defined by the recursive rule
\be{ricorsiva}
r_1=1, \qquad r_2=2, \qquad r_k=2 r_{k-2}+r_{k-1}+1. 
\ee
}

\vs

\noindent{\bf Proof.} First notice that,  if 
$\{ X_1, X_2,...,X_n\}$  
is a basis for $\cal L$,  a
neighborhood of the identity in $G$ can be obtained by varying 
$t_1,...,t_n$ in a neighborhood of the origin in $\RR^n$ in the
expression 
\be{exsr}
K:=e^{ X_1 t_1} e^{ X_2 t_2}\cdot \cdot \cdot 
e^{ X_n t_n}.  
\ee
This follows from the Inverse Function Theorem (see
e.g. \cite{sternberg} Lemma 3.1. Chp. 5 for a statement of this result) 

\vs

We now show how to generate a basis of $\cal L$ by using similarity 
transformations involving only the elements in the set $\cal S$. 

\vs 

There exist two elements $X_k,
X_i$ and a (arbitrarily small) time $\tilde t$ such that 
$e^{X_k \tilde t}X_i e^{-X_k \tilde t}$ is linearly independent from 
$ X_1,...., X_m$. If this was not the case,  we would have 
\be{contras}
e^{X_k t}X_i e^{-X_k t}=\sum_{s=1}^m a_s(t)  X_s, 
\ee
for some functions $a_s(t)$ and for every $t$. Differentiating 
(\ref{contras}) at $t=0$, we obtain 
\be{fl}
[X_k,X_i]=\sum_{s=1}^m \dot a_s(0)  X_s,
\ee
for every $k,i=1,2,...m$ which (if $dim({\cal L}) >m$) contradicts the
fact that $X_1,...,X_m$ are generators of $\cal L$. Set 
\be{XM1}
X_{m+1}:= e^{X_k \tilde t}X_i e^{-X_k \tilde t}. 
\ee 
Of course, $X_1,...,X_{m+1}$ are still a set of linearly independent
generators and therefore, as above, there exist 
two elements $ X_k,\bar
X_i$ in the set $\{X_1,...,X_m,X_{m+1}\}$
 and a (arbitrarily small) time $\tilde t$ such that 
$e^{X_k \tilde t}X_i e^{-X_k \tilde t}$ is 
linearly independent from 
$ X_1,...., X_{m+1}$. As in (\ref{XM1}), 
we obtain a new element of $\cal L$ that we call 
$X_{m+2}$, such that $\{X_1,X_2,...,X_{m+2} \}$ is an $m+2$-dimensional
subspace of $\cal L$. Proceeding this way we obtain a basis
$X_1,...,X_m,X_{m+1},...,X_n$ of the Lie algebra $\cal L$ where the
first $m$ elements are the generators we started with and the elements 
$X_{m+1},$...,$X_{n}$ are obtained via similarity transformations with
the iterative procedure we have described. Every element $X_i$,
$i=1,...,n$ can be written in the form 
\be{Xi}
X_i=e^{\bar X_rt_r}e^{\bar X_{r-1}t_{r-1}}\cdot \cdot \cdot 
e^{\bar X_1t_{1}}X_k e^{- \bar X_1t_{1}} \cdot \cdot \cdot  
e^{- \bar X_{r-1}t_{r-1}}e^{- \bar X_rt_r}, 
\ee  
where the indeterminates $\bar X_1,...,\bar X_r$ belong to $\cal S$ and the (worst
case) number of factors $r$ depends on the step we are considering. At
the first step $r=1$. At the second step the worst situation is when the
linearly independent element $X_{m+2}$ is obtained as $e^{X_i \tilde t} 
X_{m+1} e^{-X_i \tilde t}$, with $X_i \in \cal S$, in which case
$r=2$. From this point on the worst situation happens when the element 
$X_{m+k}$ is obtained as $e^{X_{m+k-2}\tilde t} X_{m+k-1}
e^{-X_{m+k-2}\tilde t}$. The number $r$ for
the $k-$th  step, $r_k$,  can be obtained by setting $r_1=1$, $r_2=2$, 
$r_k=2r_{k-2}+1+r_{k-1}$ \footnote{This is the worst case number of
factors as far as  the choice of elements used at each step is
concerned. At
the $k(>2)$-th step the worst situation happens when 
we consider $e^{X_{m+k-1}\tilde t} X_{m+k-2}e^{-X_{m+k-1}\tilde t}$ or 
 $e^{X_{m+k-2}\tilde t} X_{m+k-1}e^{-X_{m+k-2}\tilde t}$. We consider
always the second alternative which will give a lesser number of
factors.} 

\vs 

Replacing now in (\ref{exsr}) 
the expressions of $e^{X_jt}$, $j=1,...,n$  in terms of
the matrices in $\cal S$ we obtain an expression of the form 
(\ref{writtenas}). The exponentials of $X_1,....,X_m$ contribute a
single factor. The exponential of the element $X_{m+k}$, $k=1,...,n-m$ 
contributes $2r_k+1$ factors. which explains the  bound in
(\ref{bound}). 

\qed

\vs

A simple argument borrowed from the proof in \cite{lowen} 
(see also  Theorem 1.1 in \cite{CSL3} for the case $m>2$) is now
sufficient to complete the proof of the result. 

\vs 

\noindent{\bf Theorem 2.} {\it Every connected compact Lie group $G$ 
is uniformly finitely generated by any set of linearly independent 
$\{ X_1,...,X_m\}$ generators of the
corresponding Lie algebra.}  
 
\vs

\noindent{\bf Proof.} From Lemma 1, every element $K$ 
in a neighborhood $N$ of the
identity in $G$ can be expressed as the 
product of the form (\ref{key}) with indeterminates $X\in \cal S$, and
$l$ uniformly bounded. 
Now 
\be{covering}
G=\cup_{K \in G} KN.  
\ee
This is an open cover of $G$ and 
by compactness of $G$ contains a finite
 subcover. Therefore we can write 
\be{fincov}
G=\cup_{i=1}^r K_i N.  
\ee
Since from Lemma 6.2 in \cite{suss} every $K_i$ is the finite product 
of elements of the form $e^{X_i t}$, the Theorem follows. 
\qed

\vs 

\section{Extension to the Case of Nonnegative 
Parameters $t_1,...,t_l$}

From a practical point of view, it is natural to ask whether or not
every element can be expressed in the form (\ref{key}) with $l$
uniformly bounded and $t_i \geq 0$, $i=1,...,l$. In terms of
controllability of right invariant vector fields on $G$, this means that
we can reach every point in $G$ from the identity by alternately
`turning on and off' the different vector fields with a uniformly
bounded number of switches.  If we do not require the number of factors
in (\ref{key}) to be uniformly bounded the answer to this question is
already 
contained in \cite{suss} (see the remark after Theorem 6.5). We have: 

\vs

\noindent {\bf Theorem 3 \cite{suss}.} {\it 
Every element $K \in G$ can be written as 
in (\ref{key}) with $t_i \geq 0$, $i=1,2,...,l$.} 

\vs

\noindent{\bf Proof.} The elements of the form (\ref{key}), with $t_i
\geq 0$, form a semigroup $S \subseteq G$ with nonempty interior in $G$
(Lemma 6.1 in \cite{suss} and \cite{undici}) which is dense in $G$. To
prove denseness, recall that every element $K$ in $G$ can be written 
in  the form (\ref{key}) $t_i \in \RR$ and for each element $e^{-X|t|}$,
there exists a sequence of positive values $t_k \geq 0$ such that 
\be{limite}
\lim_{k \rightarrow \infty} e^{Xt_k}=e^{-X|t|}.
\ee 
In fact, pick the
sequence $e^{nX|t|}$, which by compactness has a converging subsequence 
$e^{n(k)X|t|}$. By setting $e^{Xt_k}:=e^{(n(k+1)-n(k)-1)X|t|}$ we obtain
(\ref{limite}). Since $S$ is a semigroup, with nonempty interior and
dense in $G$, it follows from Lemma 6.3 in \cite{suss} that $S=G$. \qed

\vs

A simple addition to  the argument in the previous section is needed
in order to extend Theorem 2 to the case of nonnegative parameters 
$t_i \geq 0$. First notice that from the inverse function theorem and
the proof of Lemma 1, there exist $n+1$ elements in $G$, $K_1$,...,
$K_{n+1}$, such that the image of the map $F$ from some open set in
$\RR^n_+$ to $G$, defined by 
\be{mappa}
F(t_1,t_2,...,t_n):=K_1e^{Xt_1} 
K_2e^{Xt_2}\cdot \cdot \cdot K_ne^{Xt_n}K_{n+1},  
\ee  
 is an open set in $G$.  The indeterminates 
$X$ are in $\cal S$. Using Theorem 3, we can
express $K_1$,...,$K_{n+1}$ in the form (\ref{key}) 
with $t_i \geq 0$, and therefore there exists 
an open set $U \subseteq G$ such that all 
the elements $K \in U$ can be expressed as in (\ref{key}) with $t_i \geq
0$, 
$i=1,2,...,l$, 
with a given  $l$. Therefore, using the 
same argument as in the proof of Theorem 2, we obtain the
following result. 

\vs

\noindent{\bf Theorem 4} {\it Every element 
$K$ of a connected compact Lie group  
can be written  in the form (\ref{key}) where the indeterminates $X$ are in 
a arbitrary  set $\{ X_1,...,X_m\}$ of linearly independent 
 generators of the
corresponding Lie algebra, $t_i \geq 0$, 
and the number of factors $l$ is uniformly bounded.}


\section{Concluding Remarks}

There has recently been a renewed 
interest in Lie groups decompositions
of the  form (\ref{key}) in view of their application to  
specify control laws for quantum mechanical 
systems \cite{mikoCDC} \cite{Ramaya}. Moreover, 
the original result of Jurdjevic and Sussmann has been
interpreted as a universality result for 
quantum logic gates in quantum computing \cite{LLoyd} 
\cite{NIK}. By looking at the decomposition (\ref{key}) 
as a sequence of operations on elementary 
pieces of information, one can say that 
every operation can be obtained 
by a sequence of elementary operations. In fact, 
as we have shown here, the number of
required operations is uniformly bounded.  In this case 
the relevant Lie group is the set of 
unitary matrices of appropriate dimensions.  

The paper \cite{NIK} also contains an alternative proof that every
unitary matrix can be written in the form (\ref{key}) where the
indeterminates $X$ belong to a (generic)  pair of Hermitian matrices. 
Although not explicitly mentioned by the author, this proof also works
to prove the uniform finite generation result in this special case. The
proof presented here for the general case shares some important ideas
with the one in \cite{NIK}.


\end{document}